\documentclass[final,5p,times,twocolumn]{elsarticle}
 \usepackage{graphics}
\usepackage{amssymb}
\usepackage{amsmath}

\usepackage{lineno}

\usepackage{subfigure}

\newcommand*\patchAmsMathEnvironmentForLineno[1]{%
  \expandafter\let\csname old#1\expandafter\endcsname\csname #1\endcsname
  \expandafter\let\csname oldend#1\expandafter\endcsname\csname end#1\endcsname
  \renewenvironment{#1}%
     {\linenomath\csname old#1\endcsname}%
     {\csname oldend#1\endcsname\endlinenomath}}% 
\newcommand*\patchBothAmsMathEnvironmentsForLineno[1]{%
  \patchAmsMathEnvironmentForLineno{#1}%
  \patchAmsMathEnvironmentForLineno{#1*}}%
\AtBeginDocument{%
\patchBothAmsMathEnvironmentsForLineno{equation}%
\patchBothAmsMathEnvironmentsForLineno{align}%
\patchBothAmsMathEnvironmentsForLineno{flalign}%
\patchBothAmsMathEnvironmentsForLineno{alignat}%
\patchBothAmsMathEnvironmentsForLineno{gather}%
\patchBothAmsMathEnvironmentsForLineno{multline}%
}

\journal{Nuclear Instruments and Methods A }

\begin{document}
\graphicspath{{./img/}}

\hyphenation{rec-tangular}

\begin{frontmatter}



%% SINGLE AUTHOR

\title{Alignment of the CMS Silicon Tracker -- and how to improve detectors in the future}
\author[ck]{Claus Kleinwort}
\ead{claus.kleinwort@desy.de}
\author[fm]{Frank Meier\corref{cor1}\fnref{label1}}
\ead{frank.meier@psi.ch}
\cortext[cor1]{Both authors on behalf of the CMS Collaboration}
\fntext[label1]{Corresponding author}
\address[ck]{DESY Deutsches Elektronen-Synchrotron, Notkestra\ss e 85, 22607 Hamburg, Germany}
\address[fm]{Paul Scherrer Institut, OFLC/009, 5232 Villigen, Switzerland}

\begin{abstract}
The complex system of the CMS all-silicon Tracker, with 15\,148 silicon strip and 1440 silicon pixel modules, requires sophisticated alignment procedures. In order to achieve an optimal track-parameter resolution, the position and orientation of its modules need to be determined with a precision of few micrometers. The alignment of pixels modules is crucial for the analyses requiring a precise vertex reconstruction. The aligned geometry is based on the analysis of several million reconstructed tracks recorded during the commissioning of the CMS experiment, both with cosmic rays and with the first proton-proton collisions. Statistical precision of the alignment of the module with respect to the particle trajectories to less than 10 microns has been achieved. The results have been validated by several data-driven studies (track fit self-consistency, track residuals in overlapping module regions, and track parameter resolution) and compared with predictions obtained from a detailed detector simulation.

Recent developments include the determination of sensor bow and displacements between sensors of composite modules.

Thoughts on improving future detectors with respect to alignment are given.

\end{abstract}


\end{frontmatter}

 \linenumbers

%% some new commands
\newcommand{\vekt}[1]{\mathbf{#1}}
\newcommand{\vekts}[1]{\pmb{#1}}
\newcommand{\kalman}[0]{K\'alm\'an }
\newcommand{\loremipsum}[0]{Lorem ipsum dolor sit amet, consectetur adipisicing elit, sed do eiusmod tempor incididunt ut labore et dolore magna aliqua. Ut enim ad minim veniam, quis nostrud exercitation ullamco laboris nisi ut aliquip ex ea commodo consequat. Duis aute irure dolor in reprehenderit in voluptate velit esse cillum dolore eu fugiat nulla pariatur. Excepteur sint occaecat cupidatat non proident, sunt in culpa qui officia deserunt mollit anim id est laborum.}

\section{Introduction}
\label{sec:intro}
The all-silicon inner tracker of the CMS detector at CERN consists of 15\,148 silicon strip and 1440 silicon pixel modules in a barrel-and-endacp configuration\cite{cmsjinst}. Its main purpose is to determine track parameters of charged particles produced in proton-proton and heavy ion collisions. The parameters to be delivered per track are the charge-signed curvature $\kappa$ (i.e. inverse transverse momentum), the impact parameters in the transverse plane and along the beam axis, $d_{xy}$, $d_{z}$ respectively, and the polar angles $\theta$ and $\phi$. The intrinsic hit resoultions of the detector modules are of the order of a few tens of microns, depending on module type and location. In order to determine the track parameters with high precision, the positions of the sensor modules need to be known better than their resolution. Alignment using large amounts of track data (typically several millions) is one approach to fulfill this requirement.

%-------------------------------------------------------------------------------------------------
\section{Track-based alignment}
\label{sec:aliMPII}
The use of tracks to align a tracking detector is possible under the assumption, that tracks may be described using a limited and sufficient number of parameters in an appropriate way to predict their paths. Misalignment leads to a systematic distortion of the measurements per module which can be determined using sufficiently large number of tracks and their hit signals.

In CMS, the alignment software consists of two independent algorithms, tools for the study of random and systematic misalignments and an extensive collection of tools to monitor and visualize the performance and geometry of the detector. They use track data from collision or cosmic ray muons, both as simulated and real data.

Track-based alignment relies on a suitable description of the track and its propagation through matter, as defined by the chosen detector geometry. It can be formulated as a \emph{linear least squares problem} where the following expression needs to be minimized:

\begin{equation} \label{eqn.chi2ali1} \chi^2(\vekt{p},\vekt{q}) 
	= \sum_j^\text{tracks} \sum_i^\text{hits} \vekt{r}_{ij}^T(\vekt{p},\vekt{q}_j) \,\vekt{V}_{ij}^{-1}\, \vekt{r}_{ij}(\vekt{p},\vekt{q}_j)
\end{equation}
where $\vekt{r}_{ij}$ is the residual vector containing all residuals from the tracks used and their hits, defined as 
\[\vekt{r}_{ij} = \text{track-model prediction}-\text{measured hit}.\]
The residuals are a function of $\vekt{p}$, the vector containing all alignment parameters describing the actual geometry and $\vekt{q}_j$, the track parameters of the $j^\text{th}$ track. $\vekt{V}_{ij}^{-1}$ is the inverse covariance matrix containing all information on the measurement precision and their correlations. Position and orientation of the detector modules contribute 6 or 5 degrees of freedom for silicon pixel and strip detectors, respectively. This defines the size of a subvector of $\vekt{p}$ describing one sensor.

Using a sufficiently large sample of tracks, equation \eqref{eqn.chi2ali1} and its summands follow a $\chi^2$ distribution for a corresponding \emph{number of degrees of freedom} (ndof), obeying
\begin{equation} \label{eqn.chi2probexpval}
\left\langle\frac{\chi^2(\vekt{p},\vekt{q})}{\text{ndof}}\right\rangle = 1 \qquad \left\langle\text{prob}(\chi^2,\text{ndof})\right\rangle= \tfrac{1}{2}
\end{equation}

In the case of a detector of the size as in CMS, alignment algorithms need to reduce their complexity while preserving their focus on solving the problem for two resons: 1) The total number of parameters $\vekt{p}$ and $\vekt{q}$ gets large. Aligning 16\,000 modules for position and angle leads to about 60\,000 parameters in $\vekt{p}$. The developments described in this paper will increase this number. 2) The results should be delivered within a reasonable time-frame. The two approaches in CMS to manage this are as follows:

\paragraph{Local algorithm} 
This algorithm reduces the workload by aligning the modules independently ignoring correlations at first glance. Each module is forced to the position predicted by the track hits from the other (unaligned) modules. Intrinsically, it uses the track parametrization and propagation from the CMS tracking, which takes all necessary effects of the magnetic field and material interactions into account. This approach recovers correlations between modules by iterating over the same event sample several times. At each iteration the tracks are refitted using the alignment corrections obtained in the previous iteration. Eventually the procedure yields converged alignment constants. This algorithm is known as the \emph{HIP-algorithm}\cite{arxiv0306034,CMSnote-IN2006018}.
% Veikko: The correlations between modules by design are simply taken into account by iterating over the event sample several times: at each iteration the tracks are refitted using the alignment corrections obtained in the previous iteration. Eventually the procedure yields converged alignment constants.

\paragraph{Global algorithm}
This approach reduces the complexity by the observation that it is sufficient to solve for the module parameters $\vekt{p}$ alone. This can be achieved by requiring independent measurments and the use of block-matrix theorems. This is implemented in the \emph{Millepede-II} alignment algorithm\cite{blobel06mp}. As this presentation focuses on some recent results produced using this algorithm, a more detailled description follows now.

\subsection{Millepede-II}
\label{subsec:MPII}
To accommodate for nonlinearities introduced by the track parametrization ($\vekt{q}$) and by the module parameters ($\vekt{p}$), equation \eqref{eqn.chi2ali1} needs to be linearized:

\begin{equation} \label{eqn.chi2alilin}
\chi^2(\vekt{p},\vekt{q})=
  \sum_j^\text{tracks}\sum_i^\text{hits}
  \frac{1}{\sigma_{ij}^2}
  \left(
    \vekt{m}_{ij}
    -\vekt{f}_{ij}(\vekt{p}_0,\vekt{q}_{j0})
    -\frac{\partial\vekt{f}_{ij}}{\partial\vekt{p}}\Delta\vekt{p}
    -\frac{\partial\vekt{f}_{ij}}{\partial\vekt{q}_{j}}\Delta\vekt{q}_{i}
    \right)^2
\end{equation}
where $\vekt{f}_{ij}$ is the hit position predicted by the track model from track reconstruction and $\vekt{m}_{ij}$ is the measured hit position. Assuming uncorrelated measurements allows to replace the inverse covariance matrix by $\frac{1}{\sigma_{ij}^2}$ with $\sigma_i$~the Gaussian error of the measured hit position.  

The track model used in CMS is the \kalman filter description including proper description of material effects\cite{fruehw87} and the propagation in the magnetic field\cite{strandlie06}. By design, it is a sequential fit and cannot produce the covariance matrix for all track parameters. In principle, it is possible to gain this information for all tracks a posteriori, but the \emph{Broken Lines} approach as described in \cite{blobel06bl} can be implemented more efficient and is equivalent to the \kalman approach. A brief description follows: %By construction, it accomodates for \emph{Multiple Coulomb scattering} at the detector planes.

A charged particle traversing material experiences \emph{multiple scattering}, mainly due to Coulomb interaction with the electrons in the atoms, resulting in a spatial shift and a change of the particle direction after leaving the material compared to propagation in vacuum. The mean of the deflection angle due to this effect is $\langle \beta \rangle=0$. The distribution of the deflection angles can be approximated within certain limits as a Gaussian standard deviation $\sigma(\beta)$ by the following formula\cite{pdg}:
\begin{equation} \label{eqn.mulscatt}
\sigma(\beta) = \frac{13.6\,\text{MeV}}{v p} z \sqrt{x/X_0} \left[1+0.038 \ln{(x/X_0)}\right]
\end{equation}
where $v=\beta c$ (here $\beta$ as rel.~velocity factor) is the velocity of the particle, $p$ its momentum and $z$ the charge. $x/X_0$ is the thickness of the traversed medium in units of radiation lengths.

Equation \eqref{eqn.mulscatt} takes into account all material traversed by the particle for the full trajectory. Care has to be taken during propagation, as simple summing up contributions of subpaths leads to too large estimates of $\sigma(\beta)$ by the $\ln$-term in the bracket (details in reference). It is standard procedure to treat a ``thick'' scatterer (material with a finite thickness) as two infinitely ``thin'' scatterers with same mean and sigma spaced by $1/\sqrt{3}$ of the length of the ``thick'' scatterer. In a tracking detector as in CMS, most of the material is concentrated at layers coinciding with the detector modules. They consist of matter in which the sensor interaction takes place and of non-sensing matter like support structures, cabling and cooling pipes. So the two ``thin'' scatterers coincide in the detector planes.

To determine the momentum of the charged particles' tracks, a strong and sufficiently homogeneous magnetic field of $3.8$\,T is present in the tracker. This can be taken into account by adjusting the expectation value of the scattering angle of a propagated particle $\langle\beta\rangle$ (=0 without B-field) to the value defined by the accumulated Lorentz force while propagating through the field.

Taking all this into account, the sum over all hits of one track in \eqref{eqn.chi2alilin} becomes
\begin{multline} \label{eqn.chi2bl}
\chi^2(\kappa,\vekt{u})
    = \sum_{i=1}^{n_\text{meas}}(\vekt{m}_i-\vekt{P}_i\vekt{u}_{\text{int},i})^T \vekt{V}^{-1}_{\text{meas},i} (\vekt{m}_i-\vekt{P}_i\vekt{u}_{\text{int},i}) \\
   + \sum_{i=2}^{n_\text{scat}-1} \vekts{\beta}_i(\kappa,\vekt{u})^T \vekt{V}^{-1}_{\beta,i} \vekts{\beta}_i(\kappa,\vekt{u})
\end{multline}
where $\vekts{\beta}_i$ is a vector of additional parameters of the track at every scatterer to account for the deflection angles. $\kappa$ is the charge-signed curvature, $\vekt{u}=(\vekt{u}_1,\dots \vekt{u}_{n_\text{scat}})$ describe the hit position in some local frame of the sensor and the projection matrix $\vekt{P}_i$ translates between the track frame and the local frame. The sums run over $n_\text{meas}$ recorded hits and $n_\text{scat}$ scatterers along one track, normally $n_\text{meas}<n_\text{scat}$ as the detector is neither fully hermetic nor efficient.

Solving for the minimum of eq. \eqref{eqn.chi2bl} leads to a bordered band matrix: One $\vekts{\beta}_i$ depends on the hit and its neighbours only, leading to a band matrix structure of band width $m$. The border $b$ in the matrix comes from $\kappa$, which is connected to every hit along a track. This structure allows for fast solution and determination of the covariance matrix using root-free Cholesky decomposition with a numeric complexity of $O(n^2(m+b)$, compared to $O(n^3)$ for inversion. This is needed for the refit internal to MillePede for single tracks.

All this leads to a track description equivalent to the \kalman filter model, as shown in \cite{blklmeCpc10}. It has the advantage, that the inverse covariance matrix for one track is a bordered band matrix, which can be inverted by root-free Cholesky decomposition, a faster approach than inversion.

%-------------------------------------------------------------------------------------------------
\section{More detailled surface description}
\label{sec:surfdescr}
\begin{figure}[ht] 
  \begin{center}
  \resizebox{7.5cm}{!}{\includegraphics{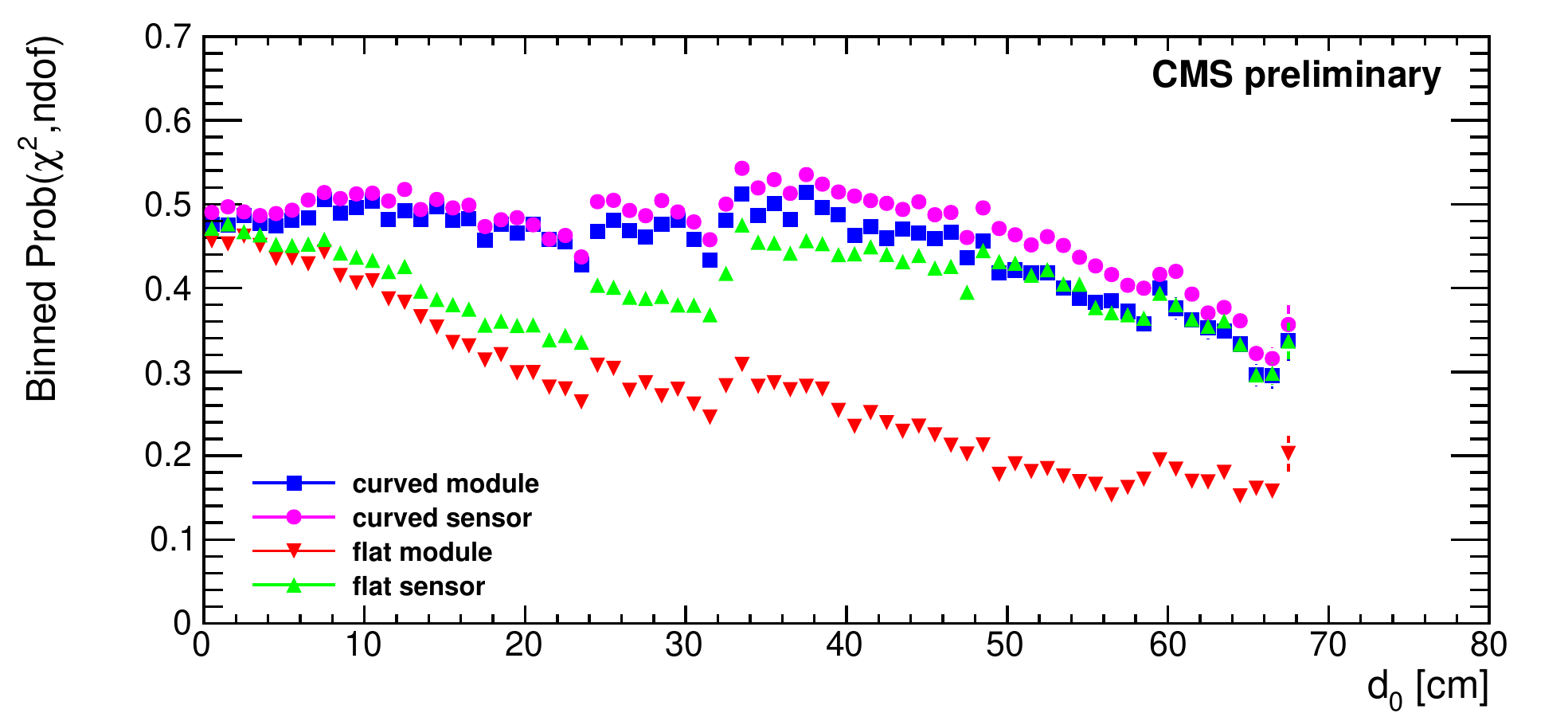}}
  \caption{\textbf{Distribution of the probability of the $\chi^2$ vs. $d_0$ (MillePede-II track refit):}
	For each track the probability of the $\chi^2$ of the MillePede internal track refit for the given number of degrees of freedom is calculated. This plot shows the average per bin, the error bars are the error of the average. The binning is done via the closest distance $d_0$ of the track to the nominal beamline. Data: Cosmic ray muons, recorded in 2009 during commissioning of CMS. 200\,000 tracks used. The results are shown for several cases with different levels of description (see text). At large $d_0 \gtrapprox 50$\,cm other effects from the track reconstruction start to dominate, which is beyond the normal use-case for the tracker and therefore neglected.}
  \label{fig:plotChi2VsD0}
  \end{center}
\end{figure}

Millepede-II uses an internal track refit as part of its procedure. Careful studies of its results as a function of track parameters have been carried out. Deviations were found while investigating tracks from cosmic ray muons. A strong dependence of the $\langle \chi^2\rangle$ on the distance of closest approach $d_{0}$ to the beamline (corresponding to track parameter $d_{xy}$) was found. This can be seen in figure \ref{fig:plotChi2VsD0}, curve for \emph{flat module}. Several hypotheses for the source of this effect have been analyzed. It is an intrinsic property of cosmic rays to have $d_0 \gg 0$ in general, which translates to a large incident angle on rectangular modules in the barrel-shaped part of the detector. For example, whenever $d_0$ is slightly smaller than the radius of a barrel layer, the track angle becomes especially large, making the hit position highly sensitive to deviations from an ideal flat rectangle. In the endcap parts of the detector, the incident angle of the cosmic ray muons is always large, independent of $d_0$.

\begin{figure*}[ht]
\centering
\subfigure[]{\label{fig:plotResidA}
\includegraphics[scale=0.4]{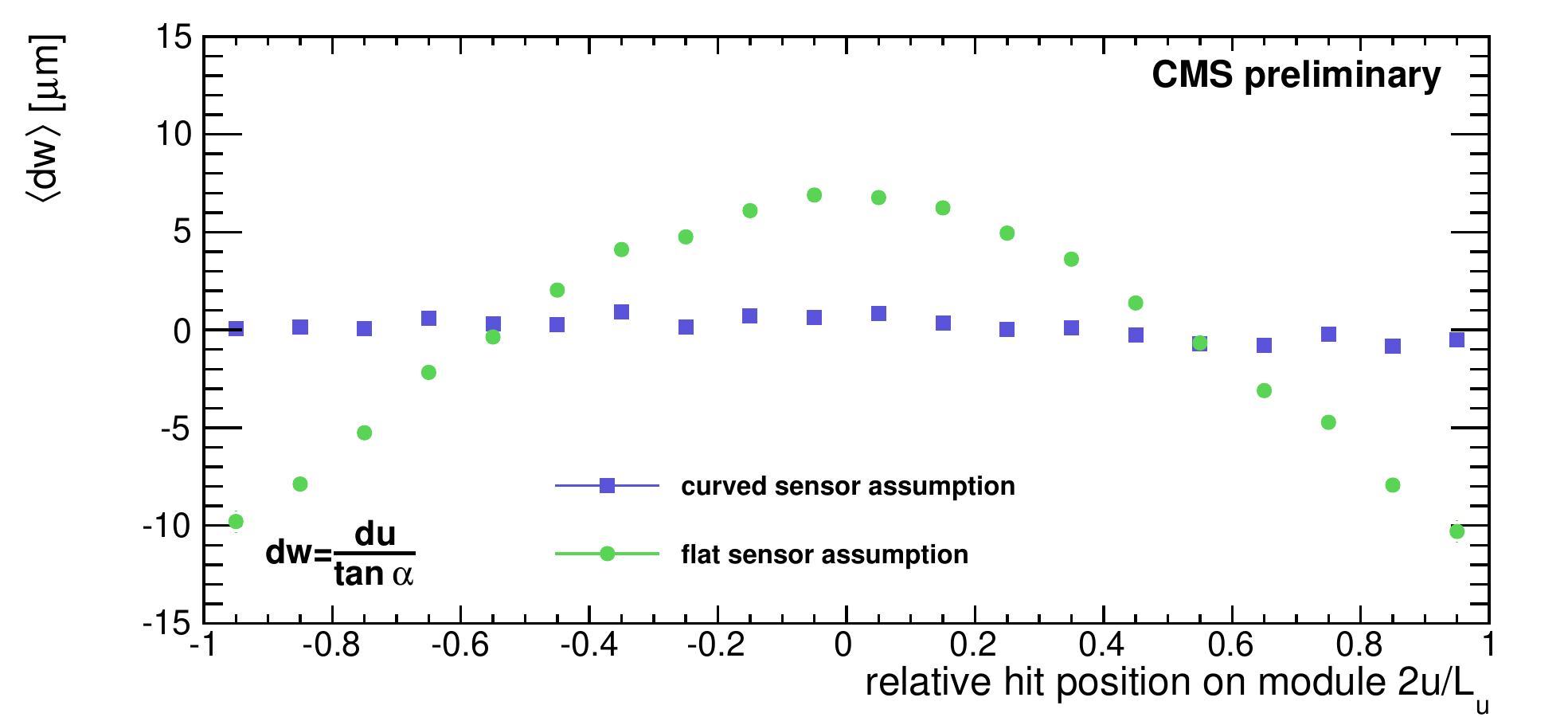}
\label{fig:untrim}
}
\subfigure[]{\label{fig:plotResidB}
\includegraphics[scale=0.4]{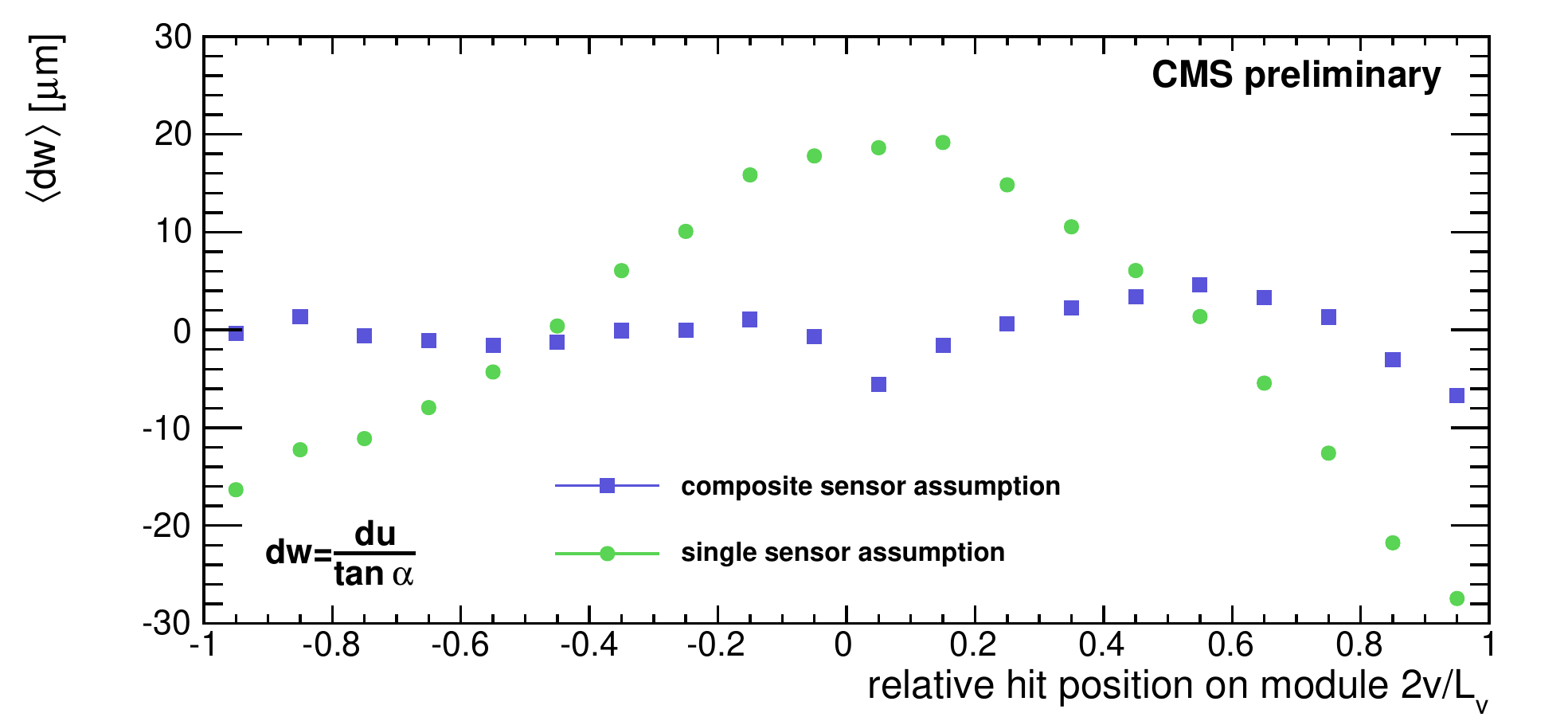}
\label{fig:trim}
}
\label{fig:plotResid}
\caption{ \textbf{Residuals perpendicular to surface along modules:}
	(a) Shown are the observed residuals in the two innermost layers of the strip barrel (tracker inner barrel, TIB), expressed as $dw=du/\tan{\alpha}$. Green circles: alignment assuming a flat surface. Blue squares: assuming a curved surface (2nd order polynomial in $u$ and $v$ plus mixed term). The measured quantities were the residual $du$ of the measured hit and the position predicted by the track fit, and the track angle $\alpha$, measured to the normal in direction of $u$. Only hits fulfilling $|\tan{\alpha}|>1/2$ have been used. Results from all modules were used, working on 200\,000 cosmic ray tracks.
	(b) Same shown along $v$ for composite modules in two innermost layers of the tracker barrel with coarser modules (tracker outer barrel, TOB). Green circles: alignment assuming one single flat sensor. Blue squares: assuming two flat sensors, splitted at $2v/L_v=0$, bows neither determined nor corrected for. Only hits fulfilling $|\tan{\alpha}|>1/2$ have been used. Results from all modules were used, working on 200\,000 cosmic ray tracks.
	The splitted surface assumption results in a flatter distribution than the single surface. Observe that in this study there was no correction for the bow, hence the right side shows a bow.  }
\end{figure*}

To take deviations from a flat surface into account, the surface has been expanded in terms of two-dimensional Legendre polynomials:
\begin{equation} \label{eqn.surfglegendre}
w(u,v)=\sum_{i=0}^N \sum_{j=0}^i c_{ij} L_j(u) L_{i-j}(v)
\end{equation}
where $w(u,v)$ is the deviation from a plane at the origin of a right-handed local cartesian coordinate system $(uvw)$ in $w$ direction as function of $u,v$. $N$ is the maximal order of the Legendre polynomials. For $N\rightarrow\infty$ every possible surface may be described. $c_{ij}$ are the orthogonal coefficients and $L_i(x)$ the Legendre polynomial of $i$-th order.

$N=1$ corresponds to the flat module assumption as used prior to this extension (except for translations in $u$,$v$ and rotations around $w$, which are not covered by \eqref{eqn.surfglegendre} and need the same treatment as before). $N=2$ introduces three additional parameters per module: $c_{20}, c_{11}$ and $c_{02}$. The first and last can be translated by choosing a proper normalization to sagittae. $c_{11}$ describes a mix-term similar to a twist. Extending the alignment to these parameters lead to the curve for \emph{flat module} in figure \ref{fig:plotChi2VsD0}. Figure \ref{fig:plotResidA} clearly shows how the residuals as a function of the hit position along strip modules recover to an expected flat distribution when taking the bowing of the sensors into account.

The sensors at larger radii of the barrel and endcaps are composite modules, i.e.\ two sensors are mounted in one module frame and daisy chained to one readout electronic block. Doing the same graph for composite modules (figure \ref{fig:plotResidB}) clearly shows another deviation from the expected curve. This is attributed to angles between the sensors. This has been implemented as well by treating the individual sensors as separate entities, which lead to the remaining curves in figure \ref{fig:plotChi2VsD0}.

Solving the alignment problem for all these added parameters lead to the determination of roughly 200\,000 parameters for the full tracker with Millepede-II in one run. This was performed on a computer equipped with a Intel Nehalem processor and 24\ GB of RAM within 6 hours of wall-clock time. Crucial parts of the algorithm were rewritten for multi-threading using OpenMP$^\text{TM}$ \cite{OpenMP} to benefit from parallel processing on the 8 cores the processor offers. The memory consumption for storing the matrix of the normal equations was reduced by using sparse matrix storage schemes and adaptive selection of storage precision of the floating-point numbers at runtime, preserving the required overall precision.

\subsection{Estimation of parameter precision}
\label{subsec:parprecestim}
\begin{figure}[ht] 
  \begin{center}
  \resizebox{7.5cm}{!}{\includegraphics{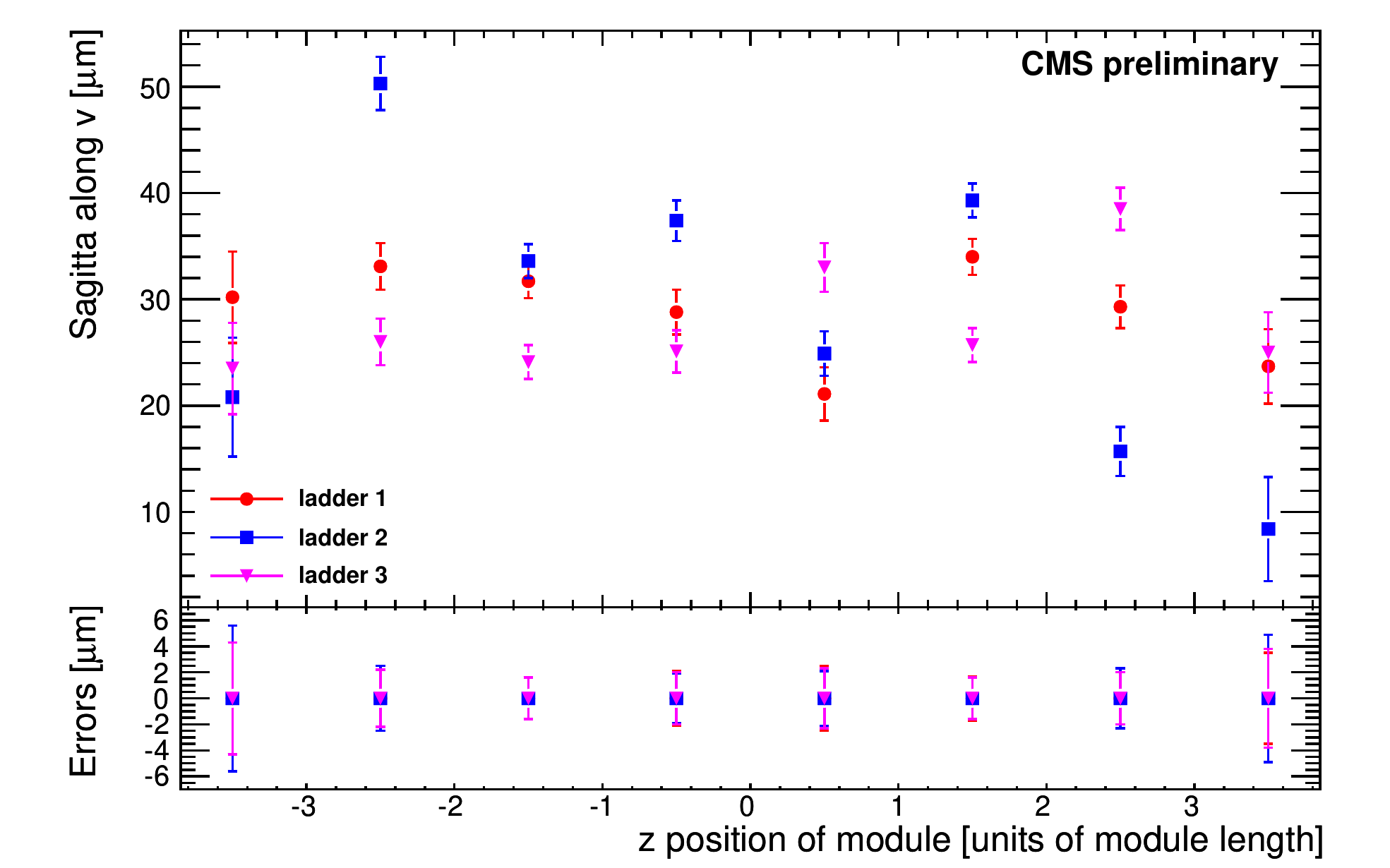}}
  \caption{\textbf{Error estimate of some parameters (statistical error only):}
	Shown are the sagittae in $v$ of pixel modules in the innermost layer, determined using a set of collision tracks (1.3 million minimum bias events at $\sqrt{s}=7\;\text{TeV}$ corresponding to 2.8 million good tracks) and cosmic ray tracks (2.5 million events, corresponding to 1.8 million tracks selected for alignment). The lower part shows the error bars centered at zero.}
  \label{fig:PXBladderSagV}
  \end{center}
\end{figure}

The Gaussian error of the parameters for the bows were estimated using the following observation: When solving for a linear least squares problem on a computer, the crucial step takes place while solving for $\vekt{x}$ in $\vekt{M}\vekt{x}=\vekt{y}$, $\vekt{M}$ being the Jacobian matrix of the normal equations. $\vekt{M}^{-1}$ would be the covariance matrix of the parameters, usually not feasible to solve for as the numerical complexity goes with $O(n^3)$ for matrix inversion, compared to other methods for solving for $\vekt{x}$. For this reason, MillePede uses the MINRES algorithm \cite{minres} as a solver instead of performing a full inversion. $\vekt{M}^{-1}$ is therefore not calculated. Individual row vectors $\vekt{M}^{-1}_i$ of $\vekt{M}^{-1}$ can be calculated by solving for $\vekt{M}\vekt{M}^{-1}_i=\vekt{\delta}_i$, where $\vekt{\delta}_i$ is the Kroneckerdelta. This has been carried out, figure \ref{fig:PXBladderSagV} shows the results for a part of the pixel barrel detector. The sagittae in the local $v$ direction can be determined to a precision of a few microns. This procedure delivers the statistical error only. No estimate on systematic errors has been carried out yet.

%\begin{table}[ht]
%  \begin{center}
%        \caption{Durations and numbers of triggers of the ROC test}
%        \label{tab:testDurations}
%%        \vspace{0.1in}
%        \begin{tabular}{l c c}
%        \hline
%                Test                         &Duration [s]  &\# triggers \\
%        \hline
%                Threshold / timing           &12            &5\\
%                Pixel test                   &1             &10\\
%                Trim bits test               &145           &5\\
%                Bump-bonding test            &80            &10\\
%                Pixel address test           &8             &1\\
%                Noise (S-Curves)             &210           &50\\
%                Trimming                     &450           &10\\
%                PH calibration               &151           &2\\
%        \hline
%        \end{tabular}
%  \end{center}
%\end{table}

%-------------------------------------------------------------------------------------------------
\section{Considerations for future detectors}
\label{sec:futuredet}
From the experience of the alignment of a tracking detector at CMS, the following considerations may be helpful in order to enhance the alignability of future detectors of similar design. \footnote{This section reflects the personal suggestions for design considerations for future detectors of one author \emph{(fm)} as given in part at the conference.} We can only speak for the configuration we know, so these thoughts need proper adjustments for other cases and are far from being universal and exhaustive.
\paragraph{Resolution} This seems to be trivial. Two aspects are worth mentioning: The fact that pixels measure two coordinates allows for alignment in all six basic degrees of freedom. Although strips can be aligned in all three rotations, doing so with pixels is far easier. A second important thing is the very high precision of the pitch along one sensor ans its constantness on the full area.
\paragraph{Module size} Larger size leads to more hits on one module for a given spatial hit density. This immediately improves the alignment precision by $\sqrt{N}$. It also improves the determination of angular alignment, as a larger module size translates to a longer lever arm.
\paragraph{Rigid mounting vs. precision mounting} Experiences from CMS show, that certain modules can be aligned even though they are displaced by few mm (sic!) from the design position. If resolution and size are already well chosen, precision mounting does not necessarily help in improving alignment. There might be some configurations for trigger layers, where precision mounting may help for other reasons.

On the other hand, rigid mounting is very important. We understand this as that the modules stay in their position over time. Track-based alignment needs data gathered over long time-periods. As it is averaging in nature, it assumes stability during the time required for recording the data it uses. Movements due to vibrations or imposed by changing conditions like temperature or magnetic field must be slower than the typical data-taking time.
\paragraph{Geometric shape} Barrel-and-endcap configurations have a great advantage for alignment: They deviate sufficiently from an ''ideal`` sphere-shaped layered detector. The modules are also flat (ore just slightly bowed), which naturally leads to a spread in incident angles on top of what the event topologies may deliver. This helps in creating constraints on several alignment modes.
\paragraph{Tracks from non-standard origin} Such tracks add more constraints on possible movements of modules which are weakly sensitive or even insensitive of changing the $\chi^2$. Cosmic ray muons are an example in the case of CMS. They come at large $d_0$ for free whereas in collision data tracks with large $d_0$ are rare events from secondary vertices. They also may connect parts of the detector with straight tracks which would normally not be connected without imposing special constraints. An example for this are the upper and lower hemispheres of a detector, which are connected with a straight track in the case of cosmic muons. Collision tracks connect these parts as well, but the use of a common-vertex constraint is necessary.
\paragraph{Optical survey} Survey has a huge drawback. Usually, survey is performed under certain artificial consitions before the final commissioning. It is a, hopefully well etsablished, assumption, that the survey data stays reliable over time.

Survey is still helpful in several ways. It delivers an independent knowledge on the geometry at the beginning of the detectors' operation. As the already mentioned tradeoff is present, the investment in survey should be limited. Think of an easy way of determining the positions of modules. In our case, position marks from the layer masks used by the manufacturing process were still visible after mounting. A standard single-reflex digital camera with a decent macro lens was used to determine relative positions of modules w.r.t their neighbours at a precision of a few micrometers. Such information can be used as an independent measurement for validation od the alignment or it may be treated as independent measurements included in the alignment algorithm.
\paragraph{Overlap} Regions with overlap are useful for alignment and monitoring of it: Particle tracks have short propagation distances and therefore their trajectories are less prone to effects imposed by multiple Coulomb scattering. The short distance between two sensors along a particle trajectory in reagions of overlap connects them together very tightly.
\paragraph{Unnecessary features} In the case of silicon detectors, implementing hardware-based alignment systems is a difficult task. Either they rely on precision mounting (e.g. some independent sensors mounted on the frame of the silicon sensors) or they mimick tracks by using lasers and holes in the metalization. Only when their precision is at least comparable to the intrinsic track hit resolution, a benefit may be realizable. They also may suffer from systematic problems, as their tracks have no geometric spread.
\paragraph{Alignment studies} The main reason why CMS acheived to align its inner tracker within that short timeframe was the use of well-known algorithms, the work of experienced people and the extensive use of a versatile alignment simulation framework. It is paramount to have the ability to simulate the detector as close to reality as possible before the final construction. There will still be surprises, like the bowed sensors.

%-------------------------------------------------------------------------------------------------
\section{Conclusions}
\label{sec:conclusion}
The inclusion of a more complex surface description of the slicon sensors of the CMS inner tracker has been shown. This was able to accommodate for discrepancies found in studies on the alignment quality and will improve the track reconstruction in CMS. The sensor bows can be determined with a statistical precision of a few micrometers. We also presented some thoughts on how future tracking detectors might benefit from the experience gathered during our work.

%-------------------------------------------------------------------------------------------------
\section{Acknowledgment}
\label{sec:acknowledgment}
The authors would like to express gratitude to Volker Blobel for the invention and implementation of the Millepe-II algorithm and the support for it.

The extensions for multithreading and storage optimizations in Millepede-II was supported by the German Helmholtz Alliance which also maintains the software package\cite{mphelmholtz}.



%% References


%\bibliographystyle{elsarticle-num}



\end{document}